# Feasibility trade-offs in decarbonisation of power sector with high coal dependence: A case of Korea

Minwoo Hyun[a], Aleh Cherp[b,c], Jessica Jewell[d,e,f], Yeong Jae Kim[g], Jiyong Eom[h*]

* Corresponding author: eomjiyong@kaist.ac.kr

[a] Department of Economics, University of California, Santa Barbara, CA, USA

[b] Department of Environmental Science and Policy, Central European University, Vienna, Austria

[c] International Institute for Industrial Environmental Economics, Lund University, Lund, Sweden

[d] Department of Space, Earth and Environment, Division of Physical Resource Theory, Chalmers University of Technology, Gothenburg, Sweden

[e] Department of Geography and Centre for Climate and Energy Transformations, Faculty of Social Sciences, University of Bergen, Bergen, Norway

[f] Risk and Resilience Program, International Institute for Applied Systems Analysis, Laxeburg, Austria

[g] RFF-CMCC European Institute on Economics and the Environment (EIEE), Centro Euro-Mediterraneo sui Cambiamenti Climatici, Italy

[h] College of Business, Korea Advanced Institute of Science and Technology (KAIST), Seoul, Republic of Korea

**Abstract**

Decarbonisation of the power sector requires feasible strategies for rapid phase-out of fossil fuels and expansion of low-carbon sources. This study develops and uses a model with an explicit account of power plant stocks to explore plausible decarbonization scenarios of the power sector in the Republic of Korea through 2050 and 2060. The results show that achieving zero emissions from the power sector by the mid-century requires either ambitious expansion of renewables backed by gas-fired generation equipped with carbon capture and storage or significant expansion of nuclear power. The first strategy implies replicating and maintaining for decades maximum growth rates of solar power achieved in leading countries and becoming an early and ambitious adopter of the CCS technology. The alternative expansion of nuclear power has historical precedents in Korea and other countries but may not be acceptable in the current political and regulatory environment.

# 1 Introduction

Many countries have embarked on profound transformations of energy systems to minimize their climate impacts while supporting economic development. Especially urgent are transitions in the power sector given its large climate impact, readily available low-carbon power generation technologies, and the importance of clean electricity for decarbonizing other economic sectors such as industry, transportation, and buildings. Many governments, therefore, have committed to eliminating or radically reducing $CO_2$ emissions from their power sector by mid-century or earlier. This requires a radical and rapid transition of electricity supply to one or several low-carbon technologies such as nuclear power, renewables, or carbon capture and storage (CCS). Are such transitions realistic in countries like the Republic of Korea that are currently heavily relying on fossil fuels in power production?

Scholars have traditionally addressed this question by modeling plausible scenarios of the evolution of power supply technologies that can, on the one hand, provide a reliable and adequate supply of electricity, and on the other hand, reduce carbon emissions to meet the targets. To support policymakers in making concrete plans, such scenarios should first contain maximum detail about future electricity systems at any given point in time. In the case of Korea, several sectoral-level decarbonisation pathways have been constructed,[1,2] but none of them included details at the level of individual power plants, which are necessary to making concrete policy decisions.

Such scenarios should also respect known constraints to the speed and scale of electricity decarbonisation. These constraints include availability of natural resources, such as hydro, solar and wind power, and land for biomass production and for storing captured $CO_2$. Other constraints include the time necessary for newer low-carbon technologies (e.g., CCS and electricity storage) to become widely commercially available, accessibility of financial resources,[3] the social acceptance of new technologies,[4] and the resistance of carbon-intensive sectors to phase-out of fossil fuels.[5]

Some of these constraints are already routinely incorporated in energy transition scenarios, while there are calls and proposals to improve underlying energy models to include as many other factors and parameters as possible.[6,7] Yet, it seems unlikely that future models will be able to incorporate all or even the most critical of constraints, because some of them are hard to quantify or generalize across countries, technologies, or time periods. Furthermore, it is often difficult to disentangle individual constraints from the aggregate effect of multiple and interacting factors, some of which may be unobservable. Hence, there have also been calls for evaluating the feasibility of energy transition scenarios based on historical experience.[8,9]

Several methods have recently been proposed for assessing the feasibility of near-term coal phase-out pledges,[10] expansion of solar and wind power,[11] and decline of fossil-powered electricity.[12] Yet most of this work has focused on assessing a particular aspect of decarbonisation at the global or continental level while the feasibility trade-offs between multiple interacting technologies in national decarbonisation scenarios have not been systematically explored. To fill the gap in the literature, a new approach to constructing and assessing the feasibility of decarbonisation scenarios is especially needed in Korea.

This paper aims to develop a set of detailed and maximally realistic electricity decarbonisation scenarios for Korea, exploring the full range of available technology options based on a fine-grained capital stock accounting model fully compatible with plant-level historical data and government projections and plans. The realism of scenarios is supported through assessing their feasibility and the trade-offs between constraints in different technology mixes in light of historical experience of energy transitions from Korea and other countries. Korea highly relies on coal and has a relatively limited hydro-, wind- and biomass power potential. At the same time, it is a large, wealthy and technologically advanced economy that is a leader in many energy technologies including nuclear power. It also has impressive national decarbonisation targets that would require ambitious power sector transformation.

The paper is constructed as follows. In the next section we provide a background to the power sector in Korea and summarize the existing approaches to feasibility. The third section contains a detailed description of our model and the scenario logic as well as the feasibility assessment method. The fourth and fifth sections report and discuss our results. In particular, we show that the major trade-off is between an ambitious built-up of nuclear energy, which is likely constrained by social acceptance, and rapid deployment of carbon capture and storage, which have uncertainties regarding technology readiness, costs, storage sites availability and unknown public attitude. The last section concludes by summarizing results and recommendations for policies and further research.

## 2  Background, theory and analytical framework

### 2.1  Background to power sector in Korea and climate and energy targets

To secure cheap and reliable electricity as a necessary means of industrial development, Korea has relied on fossil fuels and nuclear power in its power production since the 1970s for rapid economic and energy growth (Figure 1). In particular, coal and nuclear power that serve as a base load account for about 70% of total electricity supply.[13] This path frames three challenges to a rapid energy transition towards a low-carbon economy in Korea.

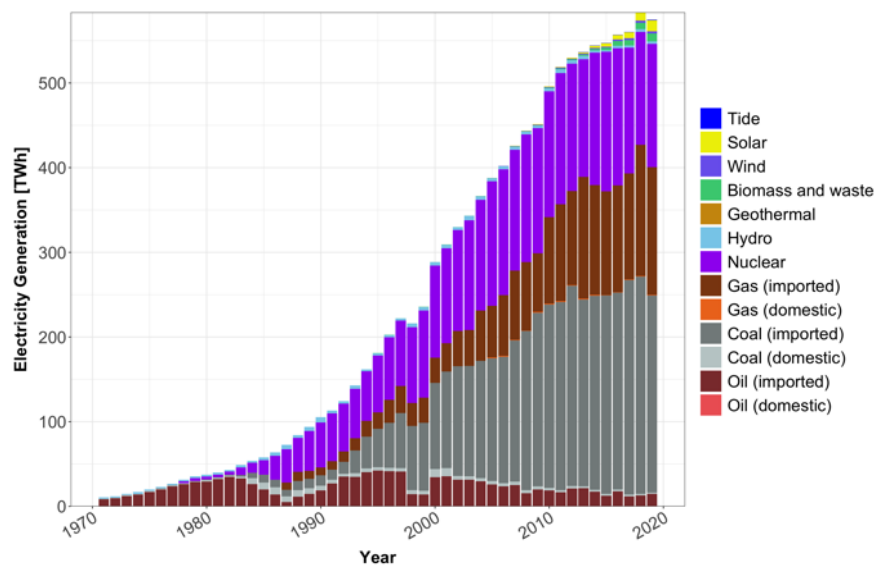

Figure 1. Electricity mix in Korea over time

First, coal power plants in Korea are on average quite young. Most of the large coal power plants (with capacity > 500MW) have been built over the last 20 years. In addition, many coal-fired plants have been granted lifetime extensions to allow performance improvement retrofits (see Supplementary Note 1). Retiring young power plants runs a risk of stranded assets and is likely to encounter stronger resistance.[10]

Second, Korea has indicated an intention to phase-out nuclear power. Korea is the fifth-largest nuclear power producer globally, and the nuclear power accounts for about one-fourth of the country's electricity generation in 2019.[14] Korea also exports nuclear technology to other countries, most recently to the United Arab Emirates (UAE). However, in recent years there have been calls to reduce the reliance on nuclear power due to the public's anti-nuclear sentiments that have emerged by several issues such as safety concerns and mistrust of the government.[15] However, simultaneous nuclear and coal phase-out like the one pursued by the 9th National Power Supply Plan[16] (NPSP) seems challenging as these are the nation's two main baseload electricity components.

Third, although the Korean power sector has experienced a steady growth of renewable capacity since 2013, the deployment of renewable sources remains limited to hydro power resources, most of which are already used and limited potential for wind production. In particular, wind power in Korea provides less than 1% of electricity generation lagging far behind most other Organisation for Economic Co-operation and Development (OECD) countries.[11] Furthermore, despite the recent growth of solar power, it still provided only 3% of the total electricity supply in 2020.[17] All renewable sources combined provided about 6% of Korea's electricity supply, the second-lowest among G20 countries. This means that expanding renewables to rapidly substitute both nuclear and coal power would represent a severe challenge.

Against these significant challenges, the Korean government is committed to accelerating the energy transition. In 2015, Korea submitted its first Nationally Determined Contributions (NDC) and presented a revised NDC roadmap in 2018, which pledged to reduce its total Greenhouse Gas (GHG) emissions by 24.4% below 2017 levels by 2030.[18] On July 14th, 2020, the government announced the Korean Green New Deal[19] to pursue a carbon-neutral society.[2] Most recently, in October 2020, the president of Korea declared to the international community its national plan to become carbon neutral by 2050.[1] According to the plan, the government will promote the rapid deployment of renewable energy (primarily solar power and wind), boosting investment in green technologies such as electricity storage systems, and CCS. The carbon neutrality target requires that Korea's power sector becomes carbon neutral or even carbon negative by 2050 as GHG emissions from electricity generation take a majority share (37.8% in 2017) of total emissions and the sector is strongly coupled with other sectors of the economy.[20]

Specifically for the power sector, the 9th NPSP was announced in December 2020, which lays out a forecast of national electricity demand, future annual plans for investment and retirement of generation units, and transmission and distribution facilities 2020 through 2034. The plan also stipulates three implementation details about the gradual phase-out of coal and nuclear units, transitioning to renewable energy sources and natural gas. First, new coal and nuclear plants will not be allowed to come online after 2024. Second, existing coal-fired units will be either retrofitted into cleaner natural gas units or pushed into early retirement. Third, existing coal-fired units will be operated at a lower utilisation rate. Fourth, renewable capacity will be increased from 20GW in 2020 to 78GW in 2034, raising the share of renewable electricity to 26.3% by 2034. These specific plant-level targets for coal phase-out and renewables expansion provide a base for our scenario assumptions to make the results more realistic and sensible.

## 2.2 Feasibility of power sector decarbonisation

Reducing carbon emissions from electricity generation requires either substituting fossil fuel combustion with low-carbon power technologies, such as nuclear or renewables, or equipping the fossil-based power plants with CCS facilities. All these solutions come with their own feasibility constraints, which are detailed in a growing body of literature.[21, 22] Our paper accounts for techno-economic, socio-technical and political contraints[23] that inform both the construction of the scenarios and the assessment of their feasibility.

Techno-economic constraints are directly incorporated in the logic of our scenarios and include supply-demand balance, infrastructure aging, and availability of natural resources.[23] Specifically, all four scenarios envision that electricity demand in Korea will steadily increase and stabilize in line with economic and population projections. Our scenarios also use a simplified relationship between the variable and non-variable sources to ensure their hour-by-hour system reliability. We further take into account the limited potential of hydro and wind power in Korea,[24] focusing on other technological solutions to decarbonisation. Finally, in considering the feasibility of CCS we refer to its costs, infrastructure requirements and geological potential for storage.

With respect to socio-technical constraints, our scenarios rely on technologies which are sufficiently mature to be rapidly expanded in Korea given national innovation systems in Korea as well as international technology diffusion.[23] Nuclear power technology has been used globally for over 60 years and in Korea for the last 50 years. Korea is one of the few advanced economies capable of maintaining, and until recently expanding, a robust domestic nuclear sector and supplying nuclear power to other countries, most recently to the UAE. On the other side, nuclear power shows signs of stagnation and decline globally[25] and there is a debate of whether its costs decrease overtime.[26, 27] To assess socio-technical feasibility of future expansion of nuclear power in Korea, we compare projected growth rates in each of the scenarios with the rates observed in Korea historically. This approach builds on the idea that historical realities are a reflection of the aggregate of causal mechanisms that will also shape the future. Such comparisons have been made globally[28, 29, 30, 31] and for individual regions[11, 32] but not for specific countries.

In comparison, the solar power sector is relatively new to Korea and began to develop only 1–2 decades ago but is steadily expanding in both Korea and around the world. Deploying solar power on a large scale would require addressing the challenge of its intermittency, to which various technological and market-based solutions are now being experimented. To assess the feasibility of rapid solar power expansion, we compare its required expansion rates with the maximum growth rates so far achieved around the world.[11] We apply a similar method to assess the feasibility of wind power deployment which is a more mature renewable power technology globally, but has experienced delayed introduction in Korea.

The third key carbon mitigation technology, CCS, is technologically ready[33] but at the moment exists primarily in demonstration plants and none in Korea except one demonstration project[1] scheduled to operate from 2025. It will arguably take more time to make CCS available and widespread in the country. To assess the feasibility of CCS deployment we compare the scale of deployment in Korea with the worldwide scale of deployment envisioned in ambitious global decarbonisation scenarios such as the ones reported by the Intergovernmental Panel on Climate Change (IPCC) Special Report1.5[34] and more recently developed by the International Energy Agency (IEA).[35]

---

[1] https://www.bloomberg.com/news/articles/2021-05-10/pumping-co2-deep-under-the-sea-could-help-korea-hit-net-zero

Finally, the expansion of low-carbon technologies may also be limited by political constraints. One type of political constraint may be the resistance of the coal sector to early retirement of coal power plants as documented by Geels et al. (2016)[5] and maybe one of the reasons for keeping countries from committing coal phase-out.[10] To parametrize this concern we assess the feasibility of coal power phase-out in Korea using the historical rate of fossil-based power decline achieved in different countries as reported by Vinichenko et al. (2021).[12] Other political constraints may limit the expansion of nuclear power or renewables. In particular, political opposition to nuclear power slowed down or stalled its expansion in many countries,[25] and it also faces strong opposition in Korea. Also, acceptance of local community has become the biggest obstacle to the diffusion of renewables, as also widely documented in other countries.[4] In our assessment, we use near- and mid-term government plans as a proxy of the political feasibility of expanding nuclear and renewables.

## 2.3 Modelling long-term decarbonisation of electricity sector

Plant-level analysis that enables the representation of existing stock and announced investment and retirement plans would be appropriate for addressing "what-if" questions regarding low-carbon energy transition. One of the valuable tools to represent long-term decarbonisation pathways in the power sector is integrated assessment models (IAMs), which typically produce a set of technology portfolios that are cost-effective in the absence of specific technology mandates or market regulations.[36] However, several concerns remain with the IAMs that endogenize technology investments, such as lack of transparency[37] and uncertainties inherent in assuming a variety of model parameters.[38]

In response to the concerns, capital stock accounting models have been employed by previous studies to improve transparency, reflect technology stock-specific policy measures, and alleviate issues associated with modeling uncertainties. The examples include models representing the stock turnover of energy-using assets such as buildings.[39, 40, 41] In addition, recent literature took a plant-by-plant accounting approach to the power sector modeling, assessing the impact on existing individual power plants (e.g., stranded assets[42] and the prioritization of retiring power plants[43]) to achieve net-zero emissions. The main advantage of this accounting approach lies in its ability to represent technology- and vintage-specific policy instruments and government plans for individual power generation units. However, little attempt has been made to assess the feasibility of power sector decarbonization scenarios based on a plant-level representation of the deployment of and substitution between alternative technology options to offer balanced insights into the required transition of the national power system.

To span realistic pathways to the power sector decarbonization for feasibility assessments, we set up a model calibrated to historical plant-level stock data. Our model has three important aspects that originate from the detailed account of individual power plants. First, scenario outcomes are easily traceable and explainable due to the simplicity and transparency of the model. For instance, when we examine the results of installed capacity by technology, it is straightforward to compare them with the historical growth of generation assets. Second, the model allows for fulfilling more precise system reliability requirements for renewables-based decarbonization. In our scenarios, the reliability requirement is benchmarked against that of the 9th NPSP. Last but not least, our modeling framework provides long-term projections that consider the construction and decommissioning plans, thereby promoting the credibility of feasibility assessment. Explicit representation of investment and retirement of major technology options based on the plant-level stock turnover model makes room for a detailed discussion of feasibility trade-offs, which is what the current study investigates.

## 3 Scenario construction

To establish a historical reference for scenario development, we collected individual plants' installed capacity, construction completion date, decommissioning date, and annual utilisation rate data spanning from 1961 to 2020 from the Korea Electric Power Corporation[2] (KEPCO). Data on country-specific emissions factors used for calculating $CO_2$ emissions was provided by the Ministry of Environment.[20] Several assumptions common across all scenarios are as follows:

- Electricity demand increases according to the prospect of the 9th NPSP—a 13.5% increase until 2034—with the growth rate gradually decreasing afterward (see Supplementary Note 2.1).
- Capacity factors for all technologies except for coal power plants are held constant, whereas capacity factors of coal power plants gradually decrease as stated in the 9th NPSP.
- Coal and gas power plants have their designed lifetime of 30 years[1] except in scenarios envisioning the early retirement of coal power plants. In scenarios with on-time and early coal retirement, no more than four coal generating units phase out simultaneously in any given year to lessen the impact on system stability. For the total 28 nuclear units, three units have their designed lifetime[2] of 30 years, 40 years for 19 units, and 60 years for the remaining units (see Supplementary Note 2.2 for more details).
- Like previous studies,[44, 45, 46, 47] gas power generation is linked to renewable power generation with the relationship governed by the system flexibility requirements implied by the 9th NPSP (see Supplementary Note 2.3).
- Oil and hydropower generation stay unchanged. As oil power plays a minor and specific niche role in power generation, it does not contribute significantly to power sector emissions. Hydropower also plays a relatively minor role in Korea and cannot be expanded due to its limited potential.
- In line with a national roadmap for CCS development,[48] CCS-installed gas power is allowed to be introduced after 2030 in CCS-containing scenarios.

To operationalize plant-level retirement and investment given the set of assumptions (for a more detailed description of scenario assumptions and variables, see Supplementary Note 2), we established a stock turnover model (Notes: Annual installed capacities ("Cap") consist of six technologies: coal, nuclear ("nuc"), natural gas ("gas"), renewables ("ren"), oil, and hydro power. The electricity demand increases by net of total investment ("Inv") and retirement ("Ret") of power plants.

Figure 2). The model has two key features. First, the model specifies total yearly installed capacity based on the age and capacity information of the individual power plants and its resulting electricity generation amount by technologies. Suppose, for example, a nuclear power unit with a nameplate capacity of 700MW and a designed lifetime of 40 years was constructed in 2000, followed by the construction of another nuclear power unit in 2010 with a 500MW capacity and 30 years. Unless new nuclear power units are allowed to be introduced, total installed nuclear capacity will decrease as much as 1,200MW in year-2040 alone. Second, new capacity investment is determined based on the residual electricity demand, which equals an increase in electricity demand net of total retirement. This plant-

---

[2] We complied with annual statistics for individual power plants presented in the KEPCO reports. They are available at the following link (in Korean):
https://home.kepco.co.kr/kepco/KO/ntcob/list.do?boardCd=BRD_000099&menuCd=FN05030103

level modeling approach makes our scenarios consistent with the existing power plant stock and the nation's technology-specific long-term investment plans.

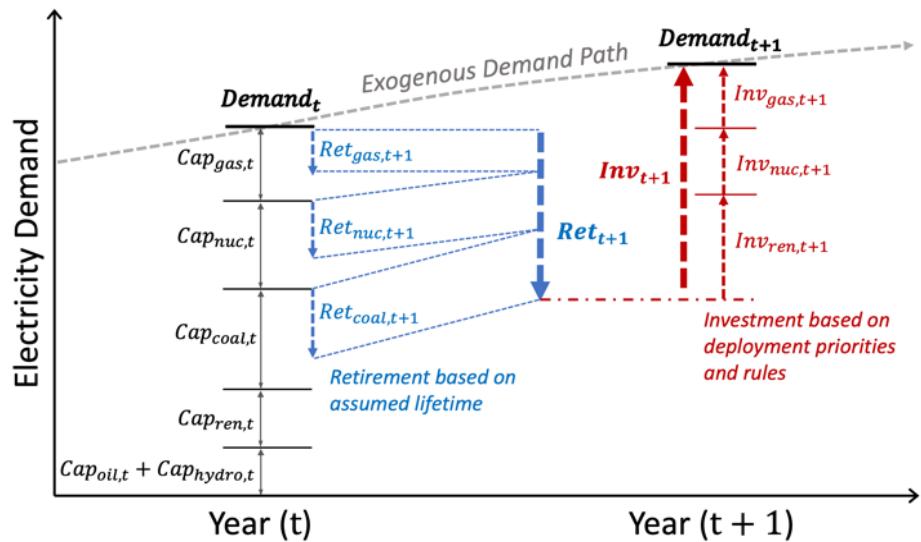

Notes: Annual installed capacities ("Cap") consist of six technologies: coal, nuclear ("nuc"), natural gas ("gas"), renewables ("ren"), oil, and hydro power. The electricity demand increases by net of total investment ("Inv") and retirement ("Ret") of power plants.

Figure 2. Technical description of stock turnover model.

In line with Korea's existing power plant stock and plans for new construction, we develop a total of ten scenarios that feature different levels of decarbonisation with the power sector's emissions ranging from 0 to 200 Mt $CO_2$ in 2050 (Figure 5). They consist of one baseline scenario (Scenario 1 in Table 1) assuming constant coal and nuclear capacities, thereby presenting no particular feasibility challenges or emission reductions, and nine policy scenarios that vary in three respects (Table 1). First, we consider four different cases depending on the degree they allow construction of new nuclear power plants: no new nuclear allowed to be introduced ("NoNuc"), current nuclear capacity held constant ("ConstNuc"), new nuclear allowed to make up for coal power retirement ("MidNuc"), new nuclear allowed to make up for coal and gas power retirement ("HiNuc"). In our scenarios, the deployment of nuclear power determines how much renewables and complementary natural gas are introduced. In NoNuc, ConstNuc, and MidNuc scenarios (Scenarios 2-8 in Table 1), the residual demand is fulfilled entirely by gas and renewables[3], with gas power deployed up to what is required for the system reliability and the remainder met by renewable power. In HiNuc scenarios (Scenarios 9-10 in Table 1), the residual electricity demand is satisfied only by nuclear and renewables without gas power. The annual increases in renewable capacity in Scenarios 9 and 10 are 1% of the total installed electricity capacity.[4] Second, the scenarios differ by whether CCS is installed for newly constructed gas units ("CCS"). Third, the scenarios differ on whether they allow for retiring coal power plants five years earlier than their 30-year lifetime to meet climate targets (suffixed as "ER").

---

[3] The renewables include solar and wind power, biomass, fuel cells, and marine power.

[4] We check whether the technology mix in the HiNuc scenarios meets the system reliability requirements. It suggests that annual flexible generation shares are within a feasible range presented in the prior studies (see Supplementary Note 2.3).

Table 1. Description of scenarios

| | Scenario | Coal power | Nuclear power | Expansion of renewables | Introduction of natural gas power with CCS |
|---|---|---|---|---|---|
| 1 | Baseline | Constant capacity | Constant capacity | To meet residual demand | None |
| 2 | ConstNuc | No new coal | Constant capacity | To meet residual demand | None |
| 3 | NoNuc | No new coal | No new nuclear | To meet residual demand | None |
| 4 | MidNuc | No new coal | Expansion to replace coal | To meet residual demand | None |
| 5 | ConstNucGasCCS | No new coal | Constant capacity | To meet residual demand | Expansion to meet flexibility requirements |
| 6 | NoNucGasCCS | No new coal | No new nuclear | To meet residual demand | Expansion to meet flexibility requirements |
| 7 | MidNucGasCCS | No new coal | Expansion to replace coal | To meet residual demand | Expansion to meet flexibility requirements |
| 8 | MidNucGasCCS_ER | No new coal & early retirement | Expansion to replace coal | To meet residual demand | Expansion to meet flexibility requirements |
| 9 | HiNuc | No new coal | Expansion to replace coal and gas | 1% of the total installed capacity per year | None |
| 10 | HiNuc_ER | No new coal & early retirement | Expansion to replace coal and gas | 1% of the total installed capacity per year | None |

# 4 Results

## 4.1 Scenario results

Our scenario results indicate that the planned phase-out of coal-fired power would necessarily require continued investments in renewables and gas power plants (Figure 3). Signficantly rapid deployment of renewable power, which amounts to an annual capacity increase of 1.5-3.0% over total system size, is to be undertaken in the short-term by 2030. In particular, scenarios with stringent limits on nuclear (NoNuc and NoNucGasCCS), which are broadly in line with the current national policy, indicate that phasing out of coal and nuclear concurrently would require unprecedentedly rapid, large-scale expansion of renewables. Note also that these scenarios would suffer from decreasing utilisation of overall generation assets due to the rapidly increasing share of renewable power (Figure 4).

Our results also indicate that in most scenarios, about six to eight 500MW-sized gas power plants have to be built every five years to fulfill the reliability requirement in response to the ramp-up of renewables. The exception is the HiNuc scenario, in which phase-out of coal and gas power together requires additions of ten to twelve 1,000MW-sized nuclear units every five years with a relatively modest increase in renewable power.

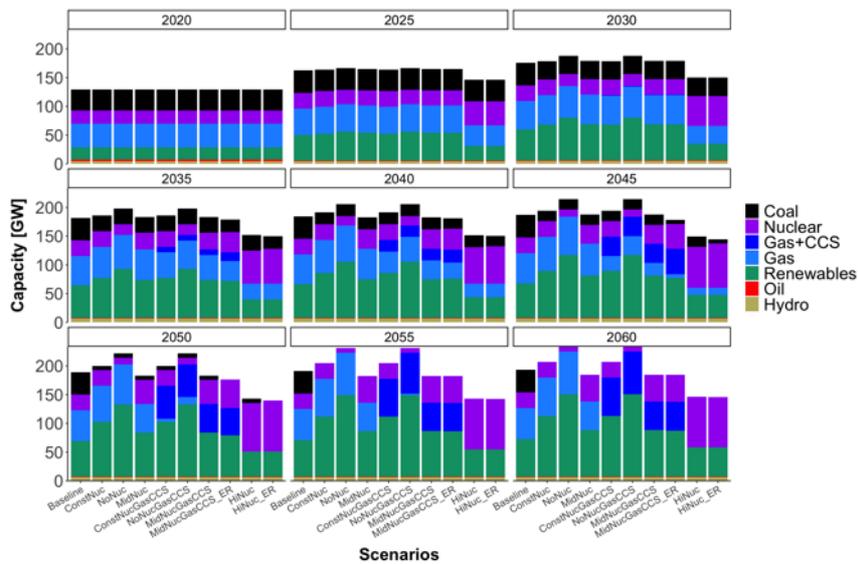

Figure 3. Installed capacity by technology in the scenarios

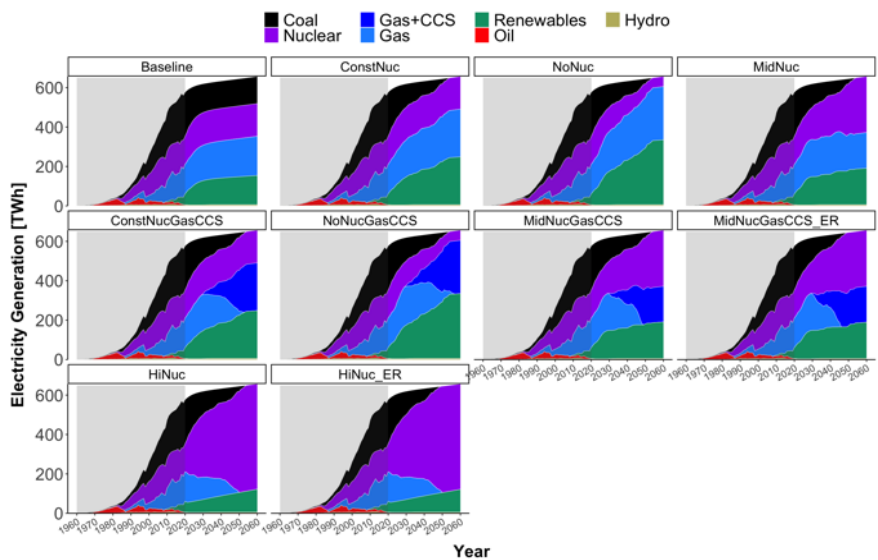

Figure 4. Annual electricity generation by technology in the scenarios

Several significant findings emerge from the comparison of $CO_2$ emissions across our scenarios (Figure 7). First, phasing out coal and expanding renewables together is not sufficient to achieve zero emissions of the power sector unless assisted by CCS or nuclear power. The three scenarios without CCS point to emissions reductions only up to about 40-50% of the current level, depending on whether nuclear is kept constant (ConstNuc), phased-out (NoNuc), or moderately expanded (MidNuc). This result is primarily due to additional $CO_2$ emissions from natural gas units, which come online to fulfill the increasing flexibility requirement of intermittent renewable power.

Second, to achieving zero emissions while phasing out nuclear power requires the rapid expansion of CCS-installed gas power in 2030-2050 (see NoNucGasCCS scenario). However, given that its near-term emissions reduction is inadequate, the 2050 zero emission target is missed by five to ten years. The inadequate emission reduction is due to the additional build-up of gas power without CCS between 2020-2030, which would not have occurred if new nuclear power had been allowed.

Third, expanding nuclear power to replace coal only (MidNuc) or coal and gas combined (HiNuc) presents more immediate, near-term emissions reductions conducive to the 2050 zero emission target than the other scenarios (Figure 5). However, without concurrent early retirement of coal power, both MidNuc and HiNuc scenarios still miss the target by about five years. Note that the allowed expansion of nuclear power combined with early coal retirement makes the on-time achievement of zero emissions possible even without CCS-installed gas power.

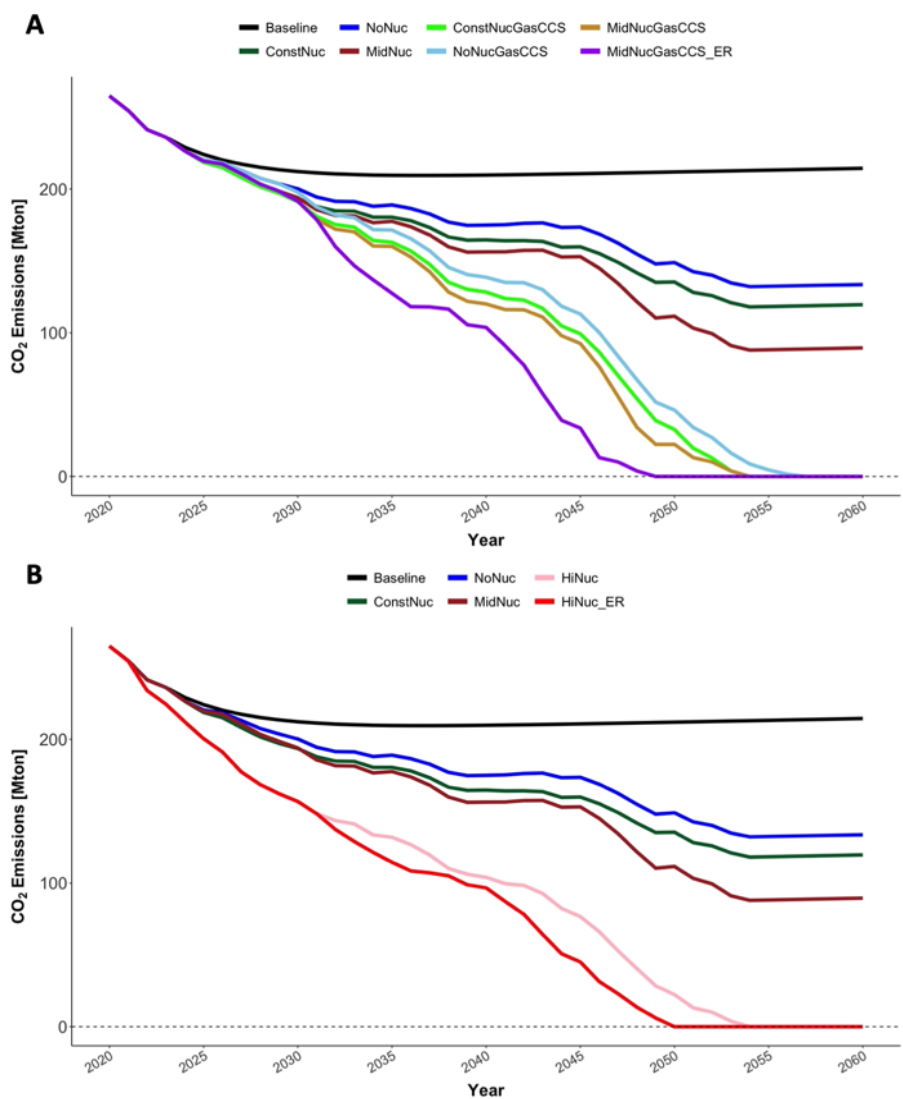

Figure 5. Annual $CO_2$ emissions in the scenarios

## 4.2 Feasibility assessments

### 4.2.1 Feasibility of solar and wind power growth in scenarios

Figure 6 illustrates the historical use of solar and wind power, national targets, and the envisioned use of renewables in our decarbonisation scenarios. The most ambitious growth of renewables is envisioned between 2020-2030, and it is much faster in no new nuclear scenarios where renewables and, to a lesser extent, gas power substitute for the rapidly declining coal generation. In comparison with the recent growth rates, the most ambitious scenarios would envision the growth of renewables at the end of the 2020s, which is about twice as fast as in 2015-2020. However, the growth of renewables so far has been accelerating, and therefore, such rates can be achieved in principle. Furthermore, similar growth has been planned in the 9th NPSP (Figure 6), which signals the existing political commitment to expand renewables with that speed, at least in the near term. However, compared to the maximum rates of renewables deployment achieved in other countries,[11] the rates of renewable expansion in Korea come across as ambitious and, on some occasions, unprecedented (Figure 7).

Another important observation is that so far, the use of renewables in Korea has been dominated by solar photovoltaic (PV). In contrast, wind power has significantly lagged behind other countries, possibly due to adverse geographic conditions, as can be observed in Japan.[49] The national plans envision the ambitious development of wind power (Figure 6), as also evidenced in recently launched projects.[5] However, there is an uncertainty of whether these adverse conditions can be overcome, which would decrease the feasibility of achieving the national targets and the more ambitious scenarios.

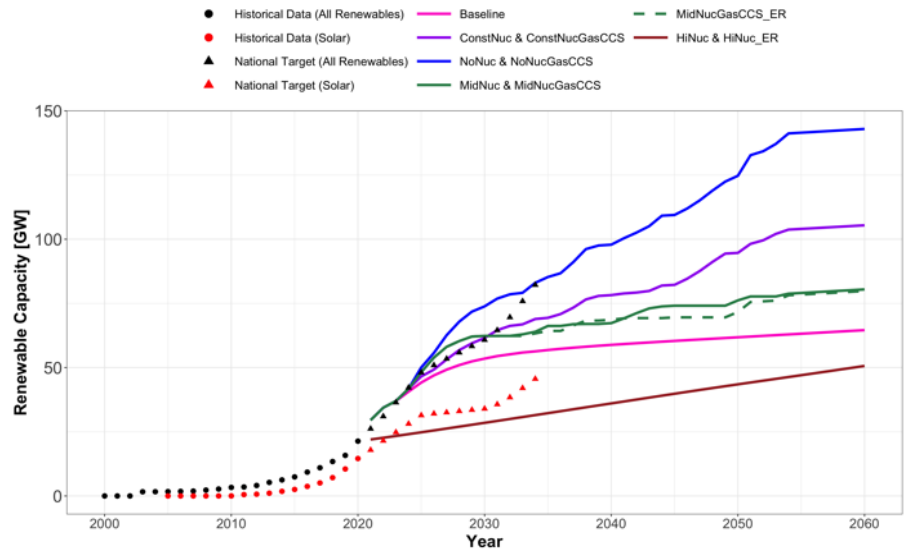

Figure 6. Historical experience, national targets and scenario results of renewables

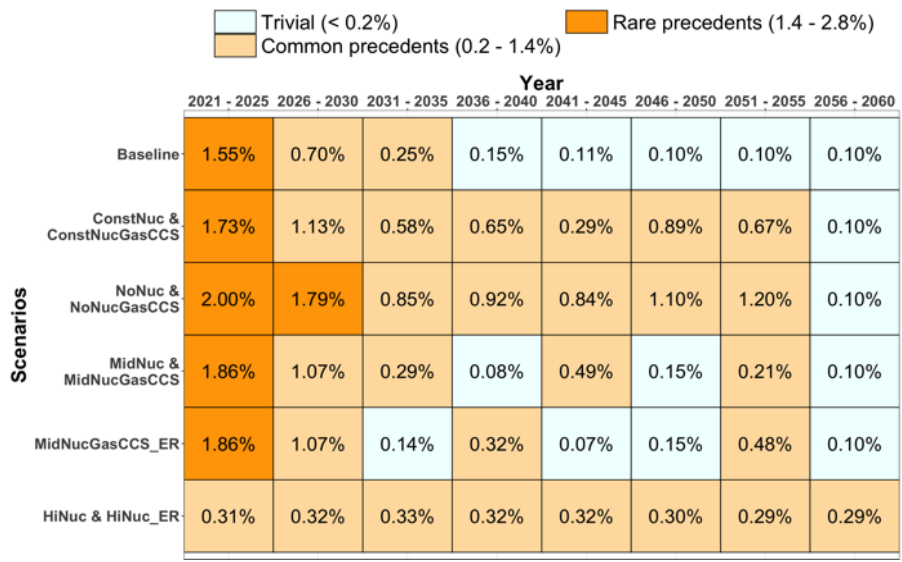

Figure 7. Growth rates of renewables in decarbonisation scenarios as compared with worldwide maximum growth rates.

---

[5] See the related article in the following link: https://www.power-technology.com/news/south-korea-wind-farm/

*4.2.2 Feasibility of nuclear power expansion*

Figure 8 shows historical trends, existing plans, and scenario projections for nuclear power, with the deployment levels and the growth rates normalized to the total electricity supply separately displayed. Historically, nuclear power in Korea experienced the fastest growth around the mid-1980s. The scenarios project different levels of nuclear power growth ranging from its gradual phase-out to the most ambitious expansion in HiNuc and HiNuc_ER. However, even in the most ambitious expansion cases, the growth is about three times slower than historically (see Supplementary Table 5 for more details).

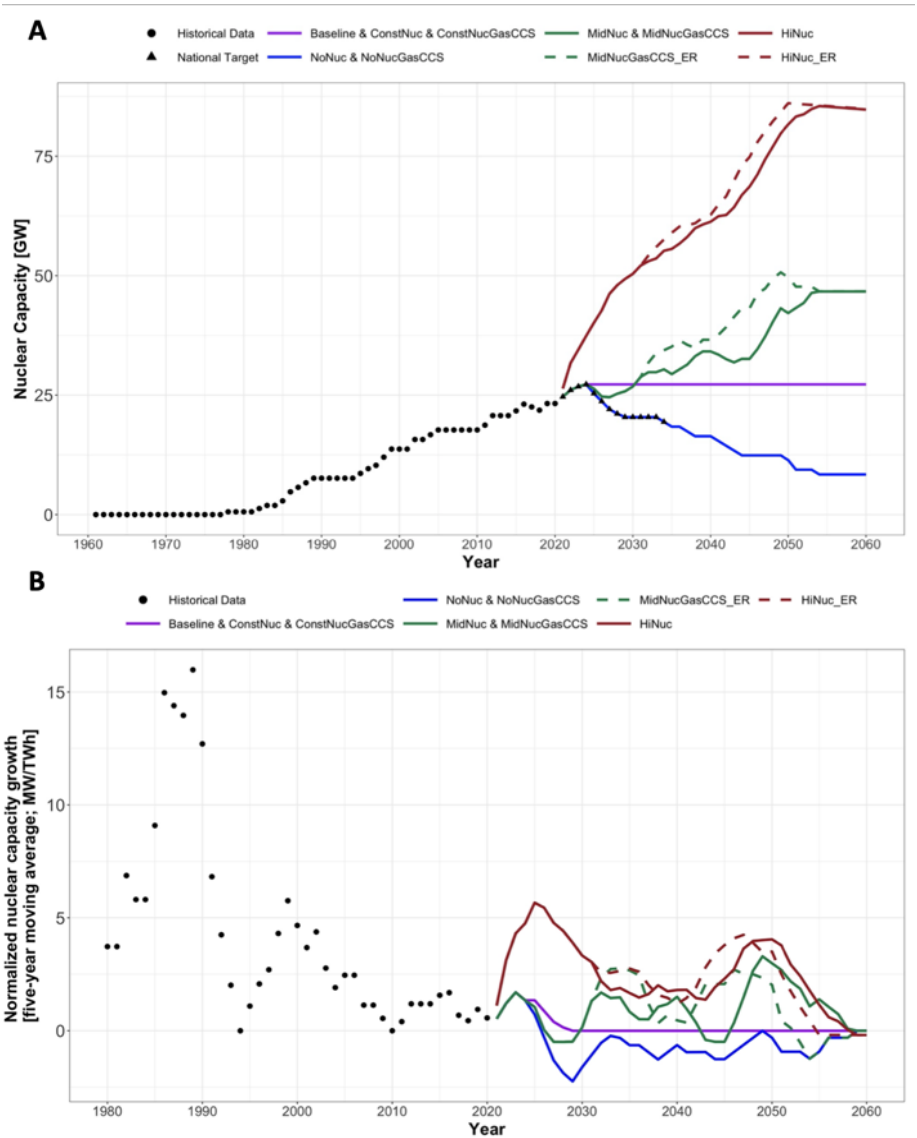

Figure 8. Historical experience and scenario results for nuclear power

Figure 9 shows the heatmap of nuclear power growth in the scenarios compared to the growth rates observed historically in different countries around the world. It should be noted that the rates of nuclear power growth in most periods and scenarios have been commonly observed historically.

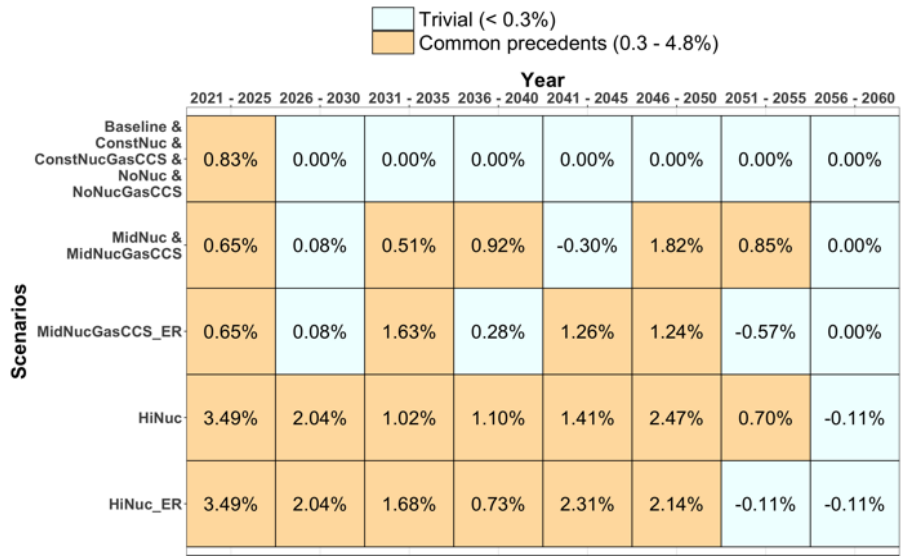

Figure 9. Nuclear power growth rates in the scenarios compared with historical rates worldwide.

### 4.2.3 *Feasibility of coal decline*

Figure 10 shows the historical growth of coal power and its future development envisioned in the scenarios, as well as the existing plans. The use of coal has rapidly increased from the 1980s onward while all scenarios, except the Baseline, indicate its equally rapid decline. Though this is a radical reversal of the national historical trends, such rapid decline has been experienced in other countries.[12] For example, in the United Kingdom (UK), coal declined by over 30 percentage points of the national electricity supply between 2007-2017. The UK case provides a realistic albeit ambitious benchmark of what would need to happen in Korea (Figure 10). While the rates of coal decline in the UK in the past and in Korea in the future would be similar, this decline in the UK has followed three decades of decline preceded by several decades of 'destabilisation',[50] while in Korea it would need to happen more rapidly. Furthermore, the average age of coal power plants in the UK at the start of its rapid decline in 2007 was 35 years,[12] while in Korea it is currently (as of 2021) 17 years.

To compare the required rates of coal power decline with a broader range of countries, we use 'feasibility zones' identified by Vinichenko et al. (2021)[12] to map the historical precedents of fossil fuel decline under different electricity demand growth rates in the same period. Figure 11 maps the decline rates projected in our scenarios onto these 'feasibility zones,' producing a heatmap similar to that of the growth of nuclear and renewables. It shows that in the early 2020s, the decline of coal has rare precedents, but subsequently, it has either multiple historical precedents or is trivial.

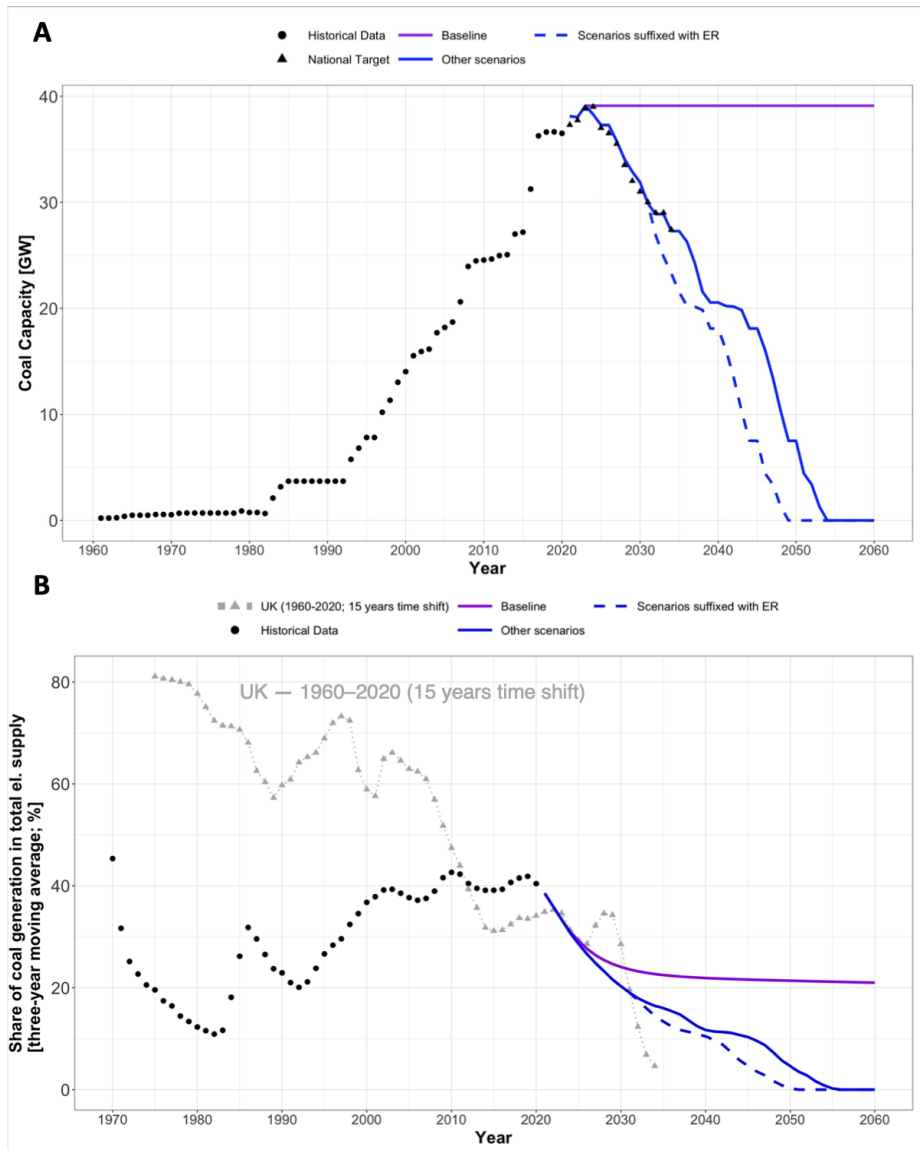

Figure 10. Historical experience and scenario results of coal power

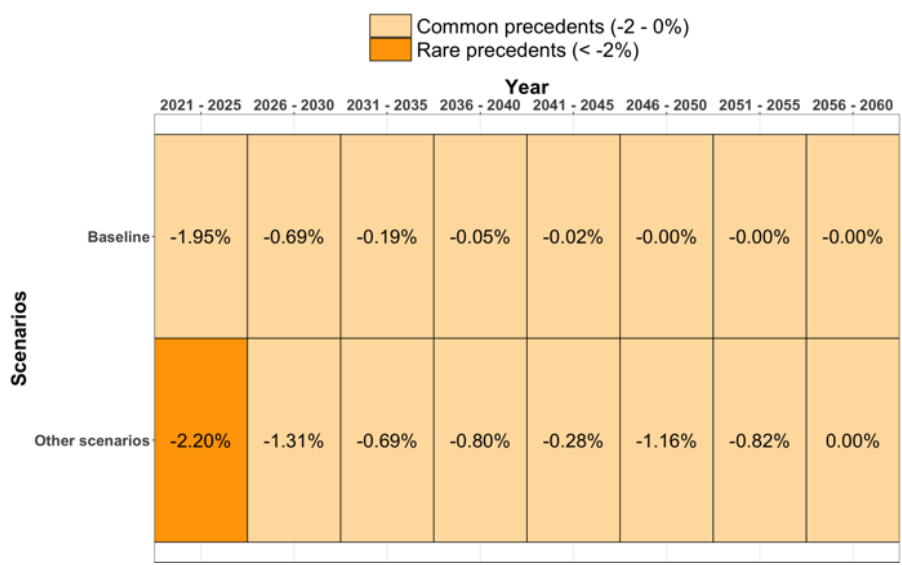

Figure 11. Rates of coal power decline in scenarios as compared to historical observed decline rates.

### 4.2.4 Feasibility of CCS

CCS is a new technology, and therefore its feasibility cannot be comprehensively evaluated based on historical experience alone. At the moment, Korea has two pilot projects with one planned project of 300 MW to be opened in mid-2020.[6] In the scenarios containing CCS, however, Korea would increase its capacity up to 70 GW capturing around 60 Mt of $CO_2$ per year by 2050. This number is more than the amount of carbon captured (20 Mt of $CO_2$) by existing CCS facilities worldwide in 2020 and about 50 times of one captured (2.4 Mt of $CO_2$) by power plants in operation.[51]

However, in the future, CCS technology is expected to develop and expand. In the IEA's central prospect, about 5.2 Gt of $CO_2$ will be captured globally by 2050, of these about 20% in the power sector (Koelbl et al. 2014[52] provide similar estimates). If the CCS-containing scenarios are realized, Korea's power sector will be responsible for some 6% of the global CCS supply while accounting for only about 1.5% of the global electricity. With respect to the CCS for gas-powered generation, Korea would need to assume even more prominent leadership, becoming responsible for up to 14% (270 TWh) of the worldwide gas power generation, which the IEA estimates to be in the range of 2000 TWh/year by 2050. This means that Korea would need to become a global leader in CCS for gas power, requiring consistent support for its research and development and deployment.

Developing CCS in Korea would require deploying its three main components: $CO_2$ capture, transportation, and storage. For **capture**, van Ewijk and McDowall (2020)[53] propose to evaluate its feasibility using flue gas desulphurization (FGD) as an analogue. Our CCS-containing scenario results show that the normalised rates of introducing CCS in Korea (7.1 – 14.4 GW/decade/Trillion USD of GDP) are comparable to the rates (11 GW/decade/Trillion USD of GDP) of historical FGD introduction (see Supplementary Note 3.2).

---

[6] The data on the pilot CCS projects completed or in development in Korea is provided by the Global CCS Institute (https://co2re.co/FacilityData).

Concerning $CO_2$ transportation and storage, although $CO_2$ is known to be storable in oil and gas reservoirs or deep saline formations, only the latter option is currently available in Korea.[54] Potential storage sites are not fully defined, but there is a possibility that both Korea and Japan might seek offshore storage. In any case, storage is most likely to be far away from capture facilities, which would mean the need to transport up to 60 Mt of $CO_2$ per year. Transportation is a mature technology, but at the moment in the whole United States (US), only 17 Mt of $CO_2$ is transported.

## 5 Discussion on the feasibility of decarbonization options

Korea's long-term energy transition pathways examined in this study differ in their emphasis on different decarbonisation options. While all envision a rapid phase-out of coal power, the essential trade-off for achieving zero emissions is between a rapid expansion of renewables backed by gas power with CCS versus expansion of nuclear power.

Table 2 summarises the feasibility of these decarbonisation options in different scenarios against the above proposed feasibility criteria. **Coal phase-out**, which is required in all decarbonisation scenarios, would be a reversal of long-term trends and unprecedented for Korea. However it is in line with the current government plans and has some international precedents, notably in the UK. The **expansion of renewables** in all scenarios except those with the rapid nuclear expansion is faster than recent national trends, but once again has some international precedents. It is also in line with the national plans and targets. Such expansion requires solar power to grow at its recent fast rates and wind power – currently lagging in Korea – to be deployed at rapid speed, as it was done in countries with more favorable geographical endowments. Although the expansion of renewables is still ambitious in the HiNuc scenario, it can be accomplished primarily by the growth of solar power with the already demonstrated speed. The **expansion of nuclear power** would be within the range of historically achieved growth rates in Korea and worldwide. However, it would contradict the recent government plans and thus may be considered politically less feasible. Finally, **the expansion of CCS**, which is especially needed in scenarios with nuclear power phase-out or stagnation, is a technologically new option with no international experience. Pursuing this option on the required large scale will likely require Korea to become one of the global pioneers in this area, which is currently not in its political commitments or plans.

Table 2. Feasibility of decarbonisation options in different scenarios

| Decarbonisation option | Scenarios | Domestic precedents | Compared to current plans | Experience of other countries | Technology readiness[7] |
|---|---|---|---|---|---|
| **Coal phase-out** | All | Unprecedented | Similar | Common precedents | N/A |
| **Expansion of renewables** | Baseline<br>ConstNuc & ConstNucGasCCS<br>NoNuc & NoNucGasCCS<br>MidNuc & MidNucGasCCS<br>MidNucGasCCS_ER | Faster than recent trends | Similar (if wind is included) | Rare precedents | Early adoption |
| | HiNuc & HiNuc_ER | In line with recent trends | Similar (if wind is excluded) | Common precedents | |
| **Nuclear power** | NoNuc & NoNucGasCCS | N/A | Similar | N/A | Mature |
| | Baseline<br>ConstNuc & ConstNucGasCCS<br>MidNuc & MidNucGasCCS<br>MidNucGasCCS_ER | Slower than past growth | More ambitious | Common precedents | |
| **CCS** | ConstNucGasCCS<br>NoNucGasCCS<br>MidNucGasCCS<br>MidNucGasCCS_ER | Unprecedented | More ambitious | Unprecedented | Demonstration |

---

[7] We follow the IEA's technology readiness level ranges from 1(concept) to 11(mature). See the related article in the following link: https://www.iea.org/articles/etp-clean-energy-technology-guide

# 6  Conclusion

This paper constructs ten scenarios of future electricity developments in Korea, of which six achieve zero emissions between 2050-2060. Analysis and comparison of these scenarios highlight the policy and practical challenges to the decarbonization of the power sector in Korea and similar countries.

First of all, all zero emission scenarios feature a complete **coal power phase-out**. The achievement of zero emissions by 2055-2060 requires that no new coal power plants are built, with none of the existing ones serving more than their prescribed lifetime of 30 years. For a more ambitious goal of achieving zero emissions by 2050, coal power should be retired even earlier. Such early retirement is in line with Korea's national carbon neutrality ambition but would be unprecedented for Korea. It would need to follow in the footsteps of the pioneers of coal retirement such as the UK and most likely deal with significant adjustments in the coal sector.

Secondly, all zero-emission scenarios envision the rapid growth of **renewable electricity**, primarily solar and wind power. The ramp-up would imply continuing the existing rapid trends of solar power deployment, initiating an equally rapid deployment of wind (that in the past experienced difficulties in taking off in Korea), and taking the electricity system into new territory featuring high penetration of intermittent renewables. Such aspirations, at least in the near term, are in line with government plans and track records of international leaders in renewable power. In scenarios that envision a rapid expansion of nuclear power, the required growth of renewables is relatively modest and can be largely achieved by maintaining the recent solar power growth rates.

The remaining challenges involve a difficult trade-off between **CCS** and **nuclear power**. On the one hand, heavy reliance on CCS would place Korea among world leaders in this new technology, which may be a challenge given its lack of fossil fuel resources and relevant technological experiences. Transportation and offshore storage of $CO_2$ may represent serious engineering challenges, raising the cost of electricity and provoking political and public opposition due to its sheer scale. While many of these challenges are uncertain and speculative in the case of CCS, the political challenges of expanding or even maintaining nuclear power seem very tangible, as clearly manifested in the government's current plans to phase out nuclear together with coal. Nevertheless, both Korea and other countries have historical experience of expanding nuclear power production at rates that, if replicated, would make CCS unnecessary. The bottom line is that at least one of these two options or their combination is absolutely necessary to achieve zero emissions from the power sector no later than 2060. Therefore, the current government plans, which envision only very modest progress on CCS and nuclear phase-out, would need to be reconsidered to face this reality.

All in all, our scenarios identify that the challenges for decarbonising the power system in Korea are formidable but manageable. This message should also give hope to other similar countries. We also propose and illustrate a new method of evaluating the feasibility of climate strategies that can be used in Korea and beyond. More work should be done to develop scenarios that can be even more informative for policymakers. Such scenarios may explore strategies for reducing uncertainties related to nuclear and CCS deployment. They can also address the twin challenge of decarbonizing the electricity sector while electrifying other economic sectors such as buildings, industry, and transportation (e.g., electric vehicles; substitution of gas appliances). In addition, the introduction of different negative emissions technologies such as bioenergy with CCS could also be taken into account.

**Acknowledgements**

This research has received funding from the European Union's Horizon 2020 research and innovation programme under grant agreement No 821471 (ENGAGE). The research by Hyun and Eom has been supported by the National Research Foundation (NRF) of Korea grants funded by the Korean government (NRF-2019K1A3A1A78112573). The authors would like to give special thanks to Hyunseok Oh for technical supports that lessen computational burdens in our scenario exercises.

# Appendix

## 1. Supplementary Discussion on Coal Power Plants

Supplementary Figure 1 depicts the installed capacity of coal by 2050 given that the assumed lifetime of the power plants is 30 years and existing units phase out as planned without any lifetime extension. Coal units with a total capacity of over 20GW have been introduced since 2000 and there is a sharp decline in capacity after the mid-2030s. Furthermore, a lifetime of most of the coal-fired plants has been extended beyond their assumed lifetime during the last 30 years (Supplementary Figure 2). It indicates that Korea has very young and large-scale coal generating units, which makes it hard to escape carbon lock-in.

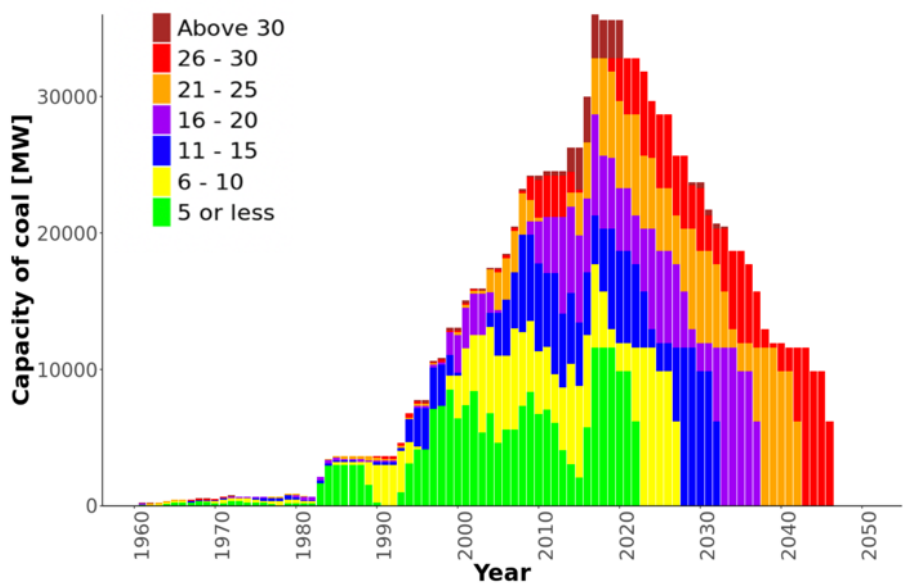

Supplementary Figure 1: Age structure of coal power plants.

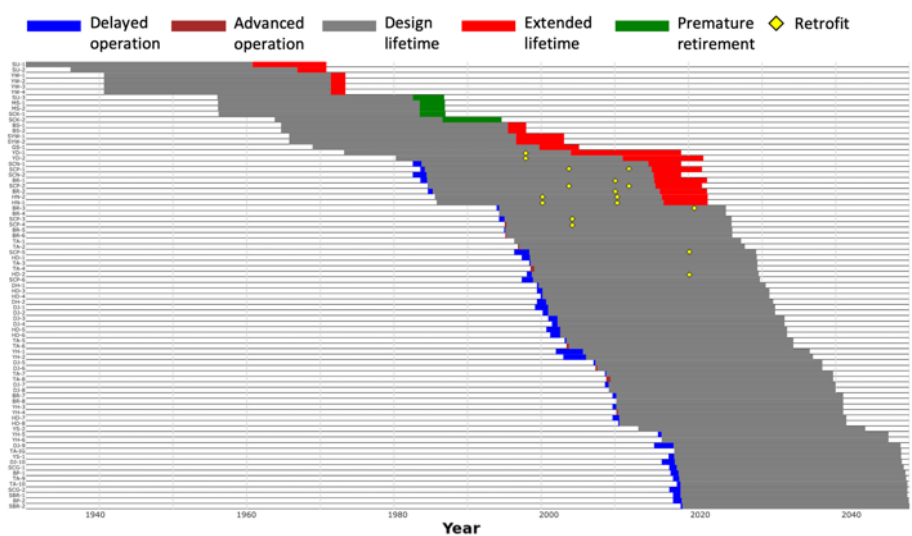

Supplementary Figure 2: Construction and decommission of individual coal power plants.

## 2. Scenario Assumptions
### 2.1. Electricity Demand

To project an electricity demand by 2060, we construct a hyperbolic decline curve (Supplementary Figure 3) for electricity demand growth rate ($GR_t$) at year $t$. The parameters of the hyperbolic function are calibrated based on the projections in the 9th NPSP (Supplementary Table 1). The equation for the growth rate is given as follows:

$$GR_t = GR_{2020} - (0.0165 \times (1 - e^{-\frac{1}{4}(t-2020)})) \tag{1}$$

Supplementary Table 1: Annual growth rates of electricity demand.

| Year | Electricity demand [GWh] | Growth rate [%] | Growth rate (9th NPSP) [%] |
|---|---|---|---|
| 2021 | 585,789 | 1.46 | 1.55 |
| 2022 | 592,695 | 1.18 | 0.58 |
| 2023 | 598,371 | 0.96 | 0.83 |
| 2024 | 603,069 | 0.79 | 0.78 |
| 2025 | 606,996 | 0.65 | 0.27 |
| 2026 | 610,311 | 0.55 | 0.41 |
| 2027 | 613,148 | 0.46 | 0.43 |
| 2028 | 615,610 | 0.40 | 0.38 |
| 2029 | 617,778 | 0.35 | 0.41 |
| 2030 | 619,715 | 0.31 | 0.56 |
| 2031 | 621,474 | 0.28 | 0.53 |
| 2032 | 623,091 | 0.26 | 0.57 |
| 2033 | 624,600 | 0.24 | 0.61 |
| 2034 | 626,025 | 0.23 | 0.56 |
| Average | | 0.58 | 0.61 |

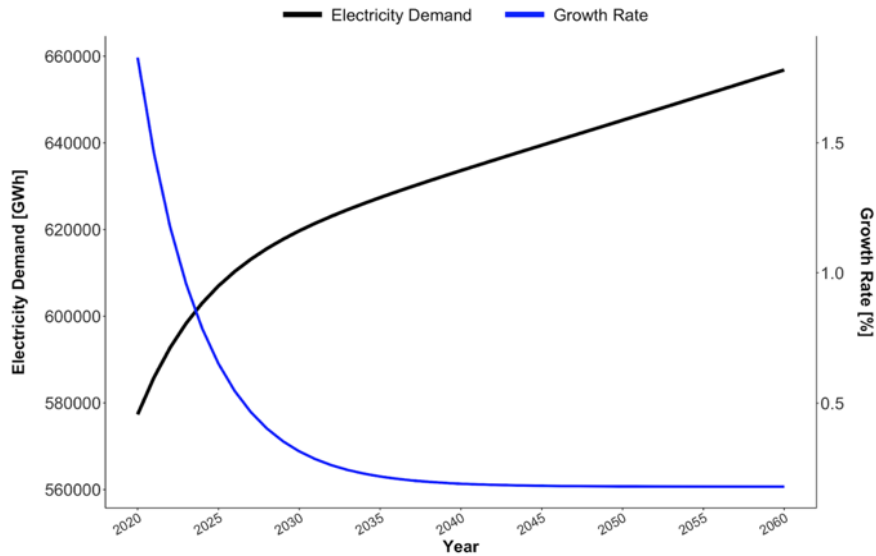

Supplementary Figure 3: Long-term electricity demand.

## 2.2. Designed Lifetimes of Nuclear Power Plants

In NoNuc and NoNucGasCCS scenarios where a new nuclear power plant is not allowed to come online, we assume that existing nuclear power plants have individual designed lifetimes detailed in Supplementary Table 2.

Supplementary Table 2: Description of individual nuclear power plants in Korea.

| Name | Unit Number | Capacity [MW] | Construction completion date | Designed lifetime [years] |
| --- | --- | --- | --- | --- |
| Kori | #2 | 650 | 1983 | 40 |
| Kori | #3 | 950 | 1985 | 40 |
| Kori | #4 | 950 | 1986 | 40 |
| Shin Kori | #1 | 1,000 | 2011 | 40 |
| Shin Kori | #2 | 1,000 | 2011 | 40 |
| Shin Kori | #3 | 1,400 | 2016 | 60 |
| Shin Kori | #4 | 1,400 | 2019 | 60 |
| Shin Kori | #5 | 1,400 | 2023 (expected) | 60 |
| Shin Kori | #6 | 1,400 | 2024 (expected) | 60 |
| Wolsong | #2 | 700 | 1997 | 30 |
| Wolsong | #3 | 700 | 1998 | 30 |
| Wolsong | #4 | 700 | 1999 | 30 |
| Shin Wolsong | #1 | 1,000 | 2012 | 40 |
| Shin Wolsong | #2 | 1,000 | 2015 | 40 |
| Hanbit | #1 | 950 | 1986 | 40 |
| Hanbit | #2 | 950 | 1987 | 40 |
| Hanbit | #3 | 1,000 | 1995 | 40 |
| Hanbit | #4 | 1,000 | 1996 | 40 |
| Hanbit | #5 | 1,000 | 2002 | 40 |
| Hanbit | #6 | 1,000 | 2002 | 40 |
| Hanul | #1 | 950 | 1988 | 40 |
| Hanul | #2 | 950 | 1989 | 40 |
| Hanul | #3 | 1,000 | 1998 | 40 |
| Hanul | #4 | 1,000 | 1999 | 40 |
| Hanul | #5 | 1,000 | 2004 | 40 |
| Hanul | #6 | 1,000 | 2005 | 40 |
| Shin Hanul | #1 | 1,400 | 2020 | 60 |
| Shin Hanul | #2 | 1,400 | 2021 | 60 |

2.3. System Flexibility Requirements

As the share of renewable generation over total electricity supply increases, it is important to incorporate system reliability constraints into the generation portfolio due to the intermittent nature of renewable power. To fulfill the system flexibility requirements, a curvilinear (quadratic) relationship is assumed between the renewables share of total generation and the flexible share of non-renewable.[44, 45] The quadratic fitting curve is obtained from the actual data presented in the 9th NPSP (Supplementary Figure 4) and its functional form is as follows:

$$y = 0.0087x^2 + 0.1584x + 11.5021 \qquad (2)$$

where $y$ denotes a flexible generation share over non-renewable generation and $x$ is a share of renewables over total electricity generation.

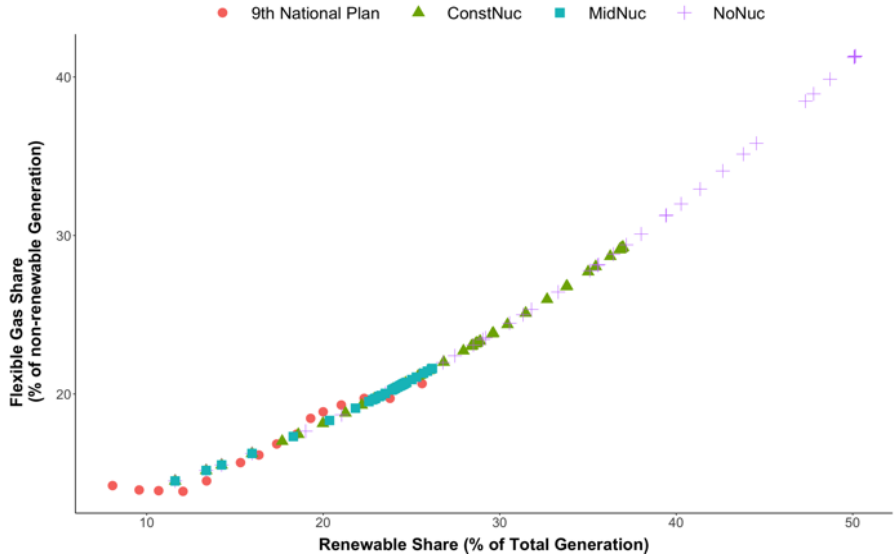

Supplementary Figure 4: System reliability requirements.

For HiNuc and HiNuc_ER scenarios, we calculate the share of electricity generation by nuclear power that is required to be utilised as a flexible generation under the flexibility constraints. Given the fixed growth rate of renewables, it shows that the flexible share of nuclear power increases over time and the maximum share is less than 18% (Supplementary Figure 5). It should be noted that previous studies[46, 55] report the potential flexible share of nuclear power as 20%, indicating that the range of the flexible nuclear power in both scenarios would be technically feasible.

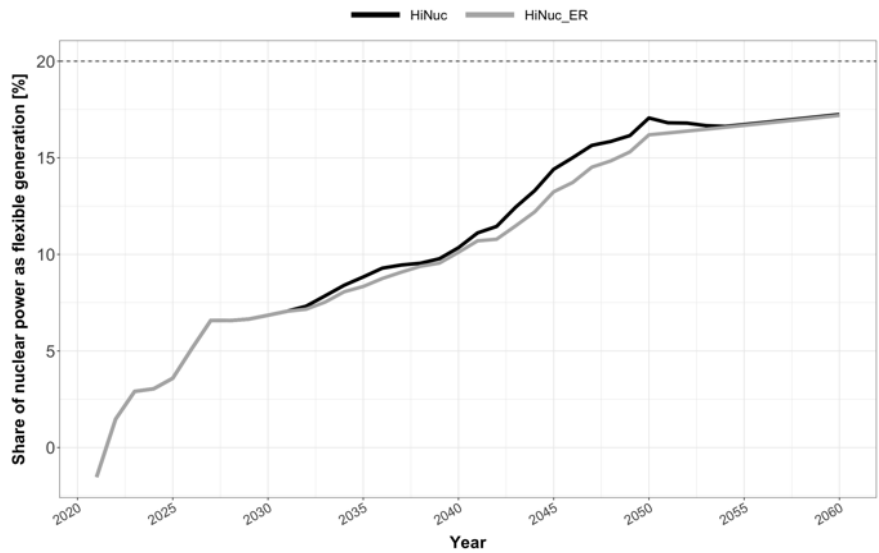

Supplementary Figure 5: Flexible nuclear share to meet system reliability requirements.

2.4. Procedures for Calculating $CO_2$ Emissions

Given the installed capacities ($CP_{i,t}$) in our stock-turnover model, we calculate the electricity generation ($Gen_{i,t}$) by technology $i$ at year $t$ using the following equation:

$$Gen_{i,t} = CP_{i,t} \times CF_{i,t} \times 8760h \tag{3}$$

where $CF_{i,t}$ is the capacity factor of the individual generation technologies. In line with the 9th NPSP, the capacity factor of coal power is decreasing from 70% in 2020 to roughly 40% in 2040 (Supplementary Figure 6), while the capacity factors for other technologies are held constant at their 2019 levels. Then, the total CO₂ emissions ($CO2_{total}$) generated from fossil fuel generation are estimated by multiplying the fuel consumption of the individual technologies by their country-specific emission factors ($EF_i$). The process can be summarized as follows:

$$CO2_{total} = \sum_i \sum_t EF_i \times Gen_{i,t} \tag{4}$$

Supplementary Table 3: Technology-specific capacity and emissions factors.

| Technology | Capacity Factor [%] (in 2019) | Country-Specific Emissions Factor [g/kWh] |
|---|---|---|
| Coal | - | 845.7 |
| Nuclear | 71.6 | 0.0 |
| Renewable | 26.3 | 0.0 |
| Gas | 41.6 | 493.7 |
| Oil | 7.6 | 661.8 |
| Hydro | 8.4 | 0.0 |

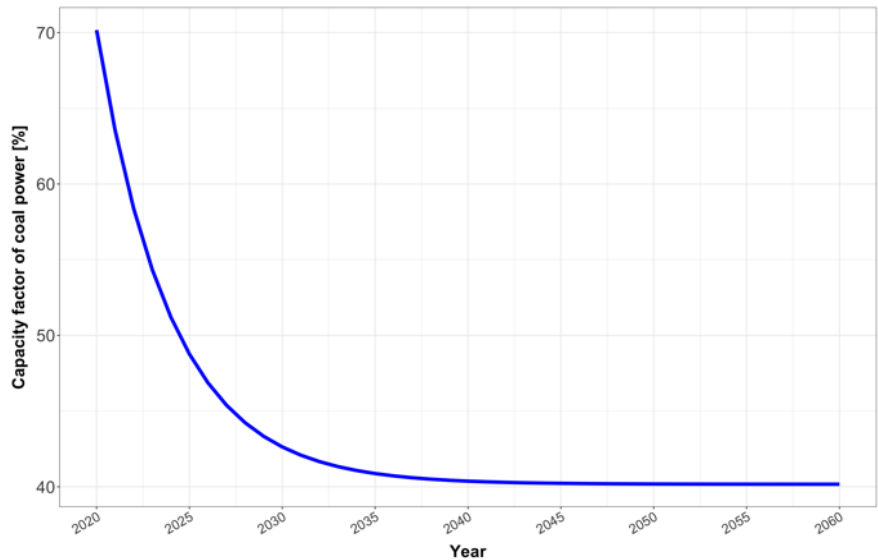

Supplementary Figure 6: Decreasing capacity factor of coal power.

## 3. Supplementary Discussion on Feasibility
3.1. Growth Rates of Renewables and Nuclear Power

Supplementary Table 4 and Supplementary Table 5 compare the highest historically observed deployment rate of renewables and nuclear power normalized to the size of the electricity system against the highest rates observed in the scenarios.

**Supplementary Table 4: Historical and scenario-wise growth rates of renewables**

| Scenario | Annual growth (MW/TWh/y) | 5-year growth (MW/TWh/5y) | Decadal growth (MW/TWh/10y) |
|---|---|---|---|
| Empirical (all renewables) | 9.6 (2020) | 24.4 (2020) | 34.2 (2020) |
| Empirical (solar) | 7.0 (2020) | 21.1 (2020) | 27.6 (2020) |
| Empirical (wind) | 0.6 (2011) | 1.6 (2019) | 3.0 (2020) |
| National Target (all renewables) | 10.1 (2034) | 41.0 (2026) | 64.7 (2034) |
| National target (solar) | 6.1 (2022) | 23.5 (2026) | 29.1 (2031) |
| National target (wind) | 3.5 (2025) | 15.5 (2027) | 28.8 (2034) |
| Scenarios (all renewables) | | | |
| Baseline | 13.9 (2021) | 41.9 (2024) | 64.5 (2027) |
| ConstNuc & ConstNucGasCCS | 13.9 (2021) | 43.8 (2024) | 72.5 (2028) |
| NoNuc & NoNucGasCCS | 13.9 (2021) | 50.7 (2028) | 92.9 (2029) |
| MidNuc & MidNucGasCCS & MidNucGasCCS_ER | 13.9 (2021) | 44.9 (2025) | 79.5 (2027) |
| HiNuc | 1.2 (2036) | 22.7 (2021) | 34.2 (2021) |
| HiNuc_ER | 1.2 (2031) | 22.7 (2021) | 34.2 (2021) |

Notes: The year for which renewables experienced the highest level of growth is reported in the parentheses.

**Supplementary Table 5: Historical and scenario-wise normalised growth rates of nuclear power**

| Scenario | Annual growth (MW/TWh/y) | 5-year growth (MW/TWh/5y) | Decadal growth (MW/TWh/10y) |
|---|---|---|---|
| Empirical (1977-2020) | 29.4 (1986) | 77.7 (1986) | 132.6 (1987) |

| Scenario | | | |
|---|---|---|---|
| Baseline & ConstNuc & ConstNucGasCCS & NoNuc & NoNucGasCCS | 2.4 (2021) | 8.5 (2023) | 11.7 (2024) |
| MidNuc & MidNucGasCCS | 4.9 (2048) | 16.5 (2049) | 22.6 (2053) |
| MidNucGasCCS_ER | 4.8 (2032) | 13.8 (2034) | 22.8 (2048) |
| HiNuc & HiNuc_ER | 9.0 (2022) | 28.2 (2025) | 44.7 (2030) |

Notes: The year for which nuclear power experienced the highest level of growth is reported in the parentheses.

### 3.2. Feasibility of CCS

A method for assessing the feasibility of the diffusion of CCS, proposed by van Ewijk and McDowall (2020),[53] is analysing a historical analogy – the diffusion of Flue Gas Desulfurization (FGD). Historical FGD capacity (black dashed line in Supplementary Figure 7) is estimated by multiplying the normalized rate of its introduction (11 GW/decade/$Trillion of GDP) by annual GDP estimates of Korea provided by the Korea Institute for Industrial Economics & Trade (2018).[56] It shows that deployment rates of gas-power with CCS in our CCS-containing scenarios are comparable to the diffusion rates based on the historical experience from FGD.

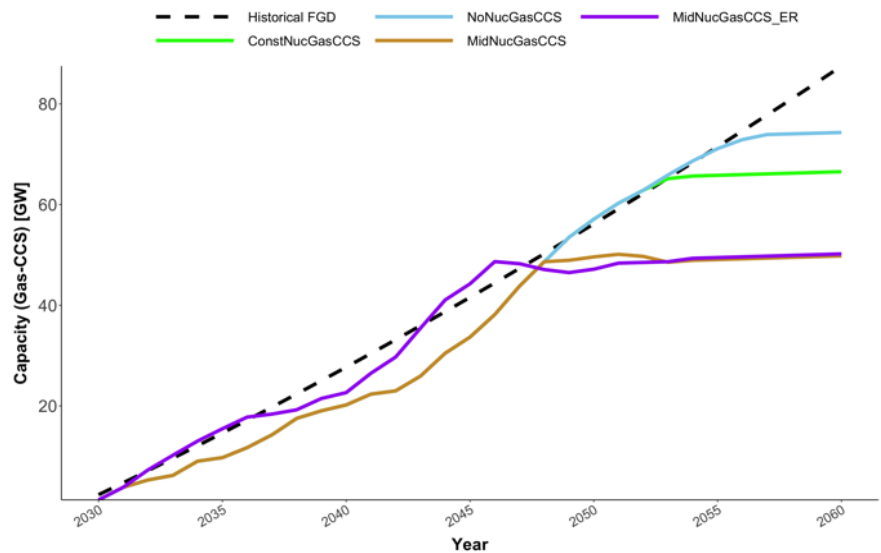

Supplementary Figure 7: Feasibility of CCS growth compared to historical FGD introduction.